\documentclass[epj]{svjour}
\usepackage{graphics}
\usepackage{amsmath}
\usepackage{amssymb}
\usepackage{color}

\begin{document}

\title{On asymptotic behavior of work distributions for driven Brownian motion}

\author{Viktor Holubec\inst{1} \and Dominik Lips\inst{2} \and Artem
  Ryabov\inst{1} \and Petr Chvosta\inst{1} \and Philipp Maass\inst{2}}

\institute{Charles University in Prague, Faculty of Mathematics and
  Physics, Department of Macromolecular Physics, V Hole{\v s}ovi{\v
    c}k{\' a}ch~2, CZ-180~00~Praha, Czech Republic
\and Universit\"at Osnabr\"uck, Fachbereich Physik,
  Barbarastra{\ss}e 7, 49076 Osnabr\"uck, Germany}

\date{Received: date / Revised version: date}

\abstract{We propose a simple conjecture for the functional form of
  the asymptotic behavior of work distributions for driven overdamped
  Brownian motion of a particle in confining potentials. This
  conjecture is motivated by the fact that these functional forms are
  independent of the velocity of the driving for all potentials and
  protocols, where explicit analytical solutions for the work
  distributions have been derived in the literature.  To test the
  conjecture, we use Brownian dynamics simulations and a recent theory
  developed by Engel and Nickelsen (EN theory), which is based on the
  contraction principle of large deviation theory.  Our tests suggest
  that the conjecture is valid for potentials with a confinement equal
  to or weaker than the parabolic one, both for equilibrium and for
  nonequilibrium distributions of the initial particle position. For
  potentials with stronger confinement, the conjecture fails and gives
  a good approximate description only for fast driving. In addition we
  obtain a new analytical solution for the asymptotic behavior of the
  work distribution for the V-potential by application of the EN
  theory, and we extend this theory to nonequilibrated initial
  particle positions.}


\maketitle

\section{Introduction}
\label{sec:introduction}

With the ever-improving possibilities to manipulate and control
physical systems on the single-molecule level, the analysis of the
role of thermodynamic quantities such as work, heat, and entropy, when
defined for single system trajectories, has become an important field
in nonequilibrium statistical mechanics
\cite{Seifert2012,Ritort2008}. One of the most intriguing achievements
in this field is the discovery of detailed and integral fluctuation
theorems
\cite{Jarzynski1997,Crooks1999,Seifert2005,Esposito2010b,VandenBroeck2010,Bochkov2013},
which in general refer to nonequilibrium systems that are externally
driven by a protocol $\lambda(t)$ of control variables during a time
interval $[0,t_f]$. These theorems hold true universally and they can
be viewed as a generalization of the second law of thermodynamics.

Among the theorems, the detailed Crooks fluctuation theorem (CFT)
\cite{Crooks1999}, and its integral counterpart, the Jarzynski
equality (JE) \cite{Jarzynski1997} are perhaps the most
prominent. They pertain to systems in contact with a heat reservoir at
temperature $T$ and in equilibrium at the initial time $t_0=0$. The
CFT states that work probability distributions (WPD) $p(w)$ and
$p_R(w)$ for a protocol $\lambda(t)$ and the associated reversed
protocol $\lambda_R(t)=\lambda(t_f-t)$, respectively, are related
according to $p(w)/p_R(-w)=\textrm{e}^{\beta (w-\Delta F)}$
\cite{Crooks1999}, where $\Delta F=F_f-F_0$ is the difference in
equilibrium free energies $F_{0,f}$ of the macrostates specified by
the control variables $\lambda_{0,f}=\lambda(t_{0,f})$, and
$\beta=(k_\mathrm{B}T)^{-1}$ is the inverse thermal energy
($k_\textrm{B}$ is the Boltzmann constant and $T$ is the temperature).
The JE states that $\langle \textrm{e}^{-\beta w}\rangle= \int
\textrm{d}w\, p(w) \textrm{e}^{-\beta w}=\textrm{e}^{-\beta \Delta F}$
\cite{Jarzynski1997}.  It becomes particularly valuable in
unidirectional experimental settings, where the work for the reversed
protocol can not be measured, and accordingly the Crooks theorem can
not be applied \cite{Palassini2011}. In applications of the JE,
measured histograms generally need to be extended to the tail regime,
because the average of $\textrm{e}^{-\beta w}$ in the JE is dominated
by rare trajectories with work values $w \ll \Delta F$. This problem
can in principle be resolved by fitting the wings of a measured WPD to
theoretical predictions for the asymptotic behavior in the limit
$w\to-\infty$.

Corresponding predictions for this asymptotic behavior are, however,
difficult to obtain. WPD depend on details of the experimental setup
and only a few generic properties have been reported so far. In the
limit of quasi-static driving, the WPD becomes a delta function,
$p(W)=\delta(W-\Delta F)$, and for sufficiently slow driving, close to
the quasi-static limit, it can be approximated by a Gaussian
distribution in the relevant regime of the integrand in the JE
\cite{Speck2004}.  The JE equality moreover gives the constraint that
$p(w) \textrm{e}^{-\beta w}$ is integrable, which implies that for
$w\to-\infty$, the WPD must decay faster than $\textrm{e}^{-\beta
  |w|}$.

For overdamped Brownian motion of a particle in time-varying
potentials, a sophisticated theory based on the contraction principle
of large deviation theory was recently developed by Engel and
Nickelsen (EN theory) \cite{Engel2009,Nickelsen2011}. This allows one
to predict the asymptotic behavior of the WPD by solving a system of
ordinary differential equations with certain boundary conditions (see
Appendix \ref{appx:ENmethod}).  Analytical solutions of these
equations were given for the harmonic potential
\cite{Engel2009,Nickelsen2011}, where in the protocol either its
minimum is moved (``sliding parabola'') or the stiffness is varied
(``breathing parabola''). In the Appendix \ref{appx:ENabsvalue} we
further derive an analytical solution for the case of a V-potential
\cite{Risken1985} with time-varying slope. However, in general, the
explicit functional form of the WPD asymptotic behavior remains
unknown and the respective differential equations have to be solved
numerically.

In this work we show that the functional form can be often guessed by
a simple method, which becomes exact in the limit of infinitely fast
driving. It is referred to as the functional form (FF) conjecture in
the following. We show in Sec.~\ref{sec:exact-results} that for the
rare cases of potentials and protocols, where the WPD has been derived
analytically, the FF conjecture always provides the correct functional
form. This leads to the question whether the FF conjecture can be
applied in general to predict the functional form of the WPD
asymptotics.

To tackle this problem, one could test the FF conjecture against
simulations. However, in corresponding simulations it becomes very
difficult to capture the WPD behavior for large negative work values
with sufficient statistics. Another possibility is to compare the
predictions of the FF conjecture with results obtained from the EN
theory. This theory, which relies on the applicability of large
deviation theory to the WPD, has been shown to reproduce the exact
Gaussian WPD for the sliding parabola \cite{Engel2009,Nickelsen2011}
and further results obtained from it for the breathing parabola turned
out to agree with an exact treatment \cite{Ryabov2013}. Moreover, the
EN theory was shown to match simulated data for the asymptotic WPD
behavior for several other potentials and protocols
\cite{Nickelsen2011,Nickelsen-thesis}.

Using the exact results for the logarithmic-harmonic potential
\cite{Ryabov2013}, we provide in Sec.~\ref{sec:exact-results} further
severe evidence of the validity of the EN theory. Given this evidence,
we then base our evaluation of the FF conjecture on the EN theory.
This evaluation is supplemented by WPD data from Brownian dynamics
(BD) simulations.  In the corresponding analysis of the FF conjecture
in Sec.~\ref{sec:equilibrium}, we will consider both ``hard
potentials'' (with stronger confinement than the harmonic one) and
``soft potentials'' (with weaker confinement than the harmonic one).
In Sec.~\ref{sec:nonequilibrium} we furthermore generalize both the EN
theory and the FF conjecture to the case of a non-equilibrated initial
position of the particle.

%%%%%%%%%%%%%%%%%%%%%%%%%%%%%%%%%%%%%%%%%%%%%%%%%
%%%%%%%%%%%%%%%%%%%%%%%%%%%%%%%%%%%%%%%%%%%%%%%%%
\section{FF conjecture and EN theory versus exact results}
\label{sec:exact-results}
%%%%%%%%%%%%%%%%%%%%%%%%%%%%%%%%%%%%%%%%%%%%%%%%%
%%%%%%%%%%%%%%%%%%%%%%%%%%%%%%%%%%%%%%%%%%%%%%%%%

%%%%%%%%%%%%%%%%%%%%%%%%%%%%%%%%%%%%%%%%%%%%%%%%%
\begin{table}
\centering
\caption{Tail behavior predicted by the FF conjecture,
  Eq.~(\ref{eq:equilibrium}), for the potentials, where the (whole)
  WPD has been calculated analytically [for certain protocols
    $\lambda(t)$, e.g.\ the form in Eq.~(\ref{eq:driving})], and for
  the V-potential (last line), where the asymptotic behavior of the
  WPD can be derived exactly (see Appendix~\ref{appx:ENabsvalue}). The
  parameters $\kappa>0$ and $g>-\beta^{-1}$ in the potential forms in
  the first and third line are constants.  The parameters $A$, $B$ and
  $C$ entering the functional form of the WPD tails are constants that
  depend on details of the protocol and the inverse temperature
  $\beta$.}
\label{tab:table1}
\begin{tabular}{l|l}
\hline\noalign{\smallskip}
\textrm{Potential $U(x,\lambda(t))$} & \textrm{Tail behavior}\\[2pt]
\hline\hline\noalign{\smallskip}\noalign{\smallskip}
$\frac{\kappa}{2} [x-\lambda(t)]^2$ &  $A{\rm e}^{-(B w - C)^2}$ \\[1ex]
$\frac{1}{2}\lambda(t) x^{2}$ & $A|w|^{-1/2} {\rm e}^{-B |w|}$\\[1ex]
$-g \log |x| + \frac{1}{2}\lambda(t) x^{2}$ 
                     & $A|w|^{-(1-\beta g)/2} {\rm e}^{-B |w|}$\\[1ex]
\hline\\[-2ex]
$\lambda(t)|x|$ &  $A{\rm e}^{-B |w|}$\\[1ex]
\noalign{\smallskip}\hline
\end{tabular}
\end{table}
%%%%%%%%%%%%%%%%%%%%%%%%%%%%%%%%%%%%%%%%%%%%%%%%%

The time evolution of the particle position $x$ for overdamped
one-dimensional Brownian motion in a time-varying potential is given
by
\begin{equation}\label{eq:langevin}
\frac{dx}{dt}=- \mu \frac{\partial U(x,\lambda(t))}{\partial x}+\eta(t)\,,
\end{equation}
where $\mu$ is the mobility, $U(x,\lambda(t))$ is the potential, and
$\eta(t)$ is a Gaussian white noise with zero mean and correlation
$\langle\eta(t)\eta(t')\rangle=2\mu k_\mathrm{B}T\delta(t-t')$. In the
following we set $k_\mathrm{B}T$ as our energy unit, and consider $x$
as dimensionless, which allows to set $(\mu k_\mathrm{B}T)^{-1}$ as
the time unit \cite{comm:units}.  The work $w$ done on the particle
along the stochastic trajectory $x(t)$ during the time interval
$[0,t_f]$ is
\begin{equation}\label{eq:work}
w=\int_0^{t_f}dt\,\frac{\partial U(x(t),\lambda(t))}{\partial\lambda}
\frac{d\lambda(t)}{dt}\,.
\end{equation}
To keep the treatment simple, we consider potentials that depend
monotonically on the protocol $\lambda(t)$ (the only exception is the
sliding parabola potential in the first line in
Tab.~\ref{tab:table1}), and we always use monotonic protocols, where
both $\partial U(x(t),\lambda(t))/\partial\lambda$ and
$d\lambda(t)/dt$ do not change sign in $[0,t_f]$.

%%%%%%%%%%%%%%%%%%%%%%%%%%%%%%%%%%%%%%%%%%%%%%%%%
\begin{figure*}
\resizebox{1.0\linewidth}{!}{%
  \includegraphics{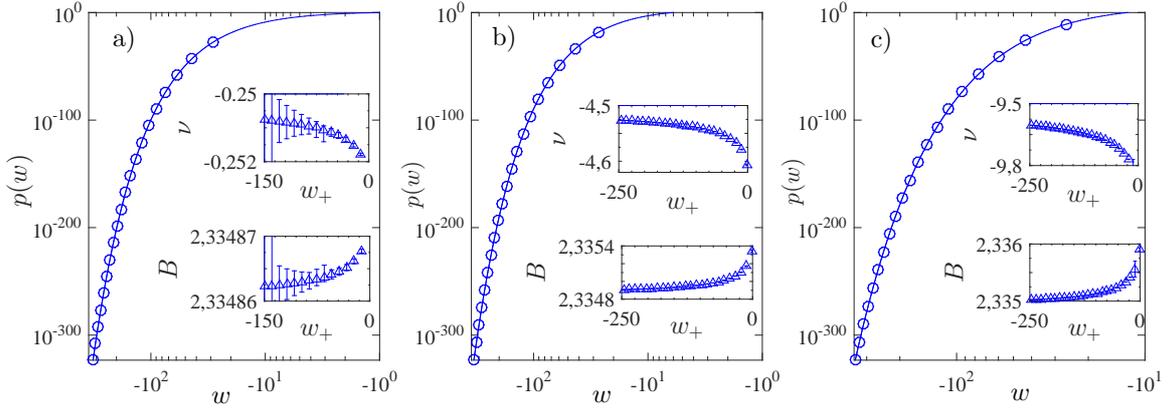}
}
\caption{Asymptotic behavior of the WPD for the logarithmic-harmonic
  potential calculated from the EN theory (symbols) and fitted to the
  asymptotic behavior $p(w)\sim |w|^{-\nu}\exp(-B|w|)$ (lines) for
  protocol parameters $r =t_f =k =1$ [see Eq.~(\ref{eq:driving})], and
  different strengths (a) $g=1.5$, (b) $g=10$, and (c) $g=20$ of the
  logarithmic part. The insets show the convergence of the fitting
  parameters $\nu$ and $B$ to their correct values $B\approx 2.335$
  (independent of $g$) and $\nu=(1-\beta g)/2$ (cf.\ third line in
  Tab.~\ref{tab:table1}) with decreasing upper bound of the fitting
  interval $[-320, w_+]$. The errors bars given in the insets, which are smaller
    than or comparable to the size of the symbols, mark 95\%
    confidence bounds of fitting parameters. The free energy differences corresponding to the 
	individual panels are (a) $\Delta F \approx -0.87$, 
	(b) $\Delta F \approx -3.81$, and (c) $\Delta F \approx -7.28$.}
\label{fig:log_harmonic_EQ}
\end{figure*}
%%%%%%%%%%%%%%%%%%%%%%%%%%%%%%%%%%%%%%%%%%%%%%%%%

In the limit of infinitely fast driving, where the protocol
$\lambda(t)$ jumps from its initial value $\lambda_0=\lambda(0)$ to
its final value $\lambda_f=\lambda(t_f)$ at some time instant in the
interval $[0,t_f]$, the WPD is given by $p(w)= \int dx\, \delta
\left[w- \Delta U(x) \right] \rho_0(x)$ where $\Delta U(x) =
U(x,\lambda(t_f)) - U(x,\lambda(0))$, and $\rho_0(x)$ is the initial
distribution of the position of the particle at time $t=0$.  In
Refs.~\cite{Nickelsen2011,Kwon2013} it was pointed out that for the
breathing parabola the \emph{functional form} of the WPD asymptotics
is the same as that for infinitely fast driving.

This suggests the following FF conjecture: If the position of the
Brownian particle is initially in equilibrium with distribution
$\rho_{\rm eq}(x) = \exp{[-\beta U(x,\lambda(0))]}/Z_0$, the
\emph{functional form} of the WPD asymptotic behavior is given by
\begin{equation}
p(w) \underset{\rm fun}{=} \int dx\, \delta \left[w- \Delta U(x)
  \right] \rho_{\rm eq}(x) \,, \quad w \!\to\! \pm\infty\,.
\label{eq:equilibrium}
\end{equation}
Here the symbol ``$\underset{\rm fun}{=}$'' indicates that the left
and the right hand-side equal in their functional form, but may differ
in specific values of parameters.  This means that in the resulting
expression after performing the integration, all coefficients
depending on $\lambda(t_f)$ or $\lambda(0)$ are considered as
unknown parameters.  For example, inserting the potential
$U(x,\lambda(t)) = \kappa[x-\lambda(t)]^2/2$ from the first line of
Tab.~1 into Eq.~(\ref{eq:equilibrium}), one obtains $p(w)\underset{\rm
  fun}{=}
\exp[-\beta(2w-\kappa\Delta\lambda^2)^2/(8\kappa\Delta\lambda^2)]/|\kappa
\Delta\lambda|$ after integration, where $\Delta\lambda = \lambda(t_f)
- \lambda(0)$. The constants containing $\Delta\lambda$ are considered
as unknown parameters and marked by $A$, $B$ and $C$ in the table,
giving $p(w)\underset{\rm fun}{=} A\exp[-(Bw-C)^2]$. Another example
is the potential $U(x,\lambda(t)) = \lambda(t) x^2/2$ from the second
line of Tab.~1. In this case one obtains from
Eq.~(\ref{eq:equilibrium}) $p(w)\underset{\rm fun}{=}|\Delta \lambda
w|^{-1/2}\exp[-\beta\lambda(0)w/\Delta \lambda]$. The resulting
functional form is $p(w)\underset{\rm fun}{=}A|w|^{-1/2}\exp[-B |w|]$,
because the exponent $-1/2$ does not depend on the driving $\Delta
\lambda$.

The first three potentials in Tab.~\ref{tab:table1} refer to those,
where exact analytical results for the complete WPD have been derived,
namely (i) the sliding parabola \cite{Mazonka1999}, (ii) the breathing
parabola \cite{Speck2011,Ryabov2013}, and (iii) the
logarithmic-harmonic potential \cite{Ryabov2013}. As a fourth example
we added the V-potential, where the asymptotic behavior predicted by
the EN theory could be derived analytically (see
Appendix~\ref{appx:ENabsvalue}). For all these potentials, where the
asymptotic behavior of the WPD is known analytically, the FF
conjecture on the functional form is correct.

Let us note that in Eq.~(\ref{eq:equilibrium}) we assumed that $\Delta
U(x)$ is different from zero, which always is the case for a monotonic
protocol ($d\lambda(t)/dt$ strictly positive or negative in
$[0,t_f]$).  Moreover, in Eq.~(\ref{eq:equilibrium}) we not only refer
to the asymptotic behavior for $w\to-\infty$ (relevant for the JE),
but also to the limit $w\to+\infty$. The WPD asymptotic behavior in
the latter limit is also in agreement with all analytically known
results. The surprising agreement suggests that the functional form of
WPD asymptotic behavior does not depend on the velocity of the driving
and gives us the motivation to further test the FF conjecture.

The EN theory is exact in the weak noise limit, i.e., up to the first
order in thermal energy $k_BT$ [see Eq.~~(\ref{eq:tails_Engel}) in
  Appendix~\ref{appx:ENmethod}]. The thermal energy determines the
free diffusion coefficient $D_0 \propto k_BT$ and thus the weak noise
regime means that the internal dynamics of the system is slow. For
driven systems, the slowness should be compared with the speed of the
driving. In this respect, a slowly driven system can be obtained by
either fixing the driving speed and increasing the temperature or vice
versa.

As discussed in the Introduction, we first provide further compelling
evidence that EN theory is indeed valid also for slowly driven
systems.  We find this necessary, because (i) confirmation of this
theory by analytical results has so far been done only for the
harmonic potential, where the propagator for the dynamics is Gaussian,
and (ii) the EN theory for other potentials has been tested by BD
simulations, which, however, are restricted to relatively fast
driving, where the tails of the WPD can be determined with sufficient
statistics.  For slow driving, the essential part of the WPD is always
well described by a Gaussian distribution \cite{Speck2004}, while the
asymptotic behavior shows up only when going to extremely large $|w|$
values.

The compelling evidence is obtained by utilizing the exact analytical
results for the logarithmic-harmonic potential $U(x,\lambda(t))=- g \log
(|x|) + \lambda(t) x^{2}/2$ \cite{Ryabov2013}, see third line in
Tab.~\ref{tab:table1}, because in the presence of the logarithmic part
($g\ne0$) the propagator for the dynamics is no longer Gaussian, and
the pre-exponential factor depends sensitively on the strength $g$ of
the logarithmic part and the inverse temperature $\beta$. For slow
driving, the asymptotic regime is expected to occur at very large
negative work values with extremely small $p(w)$ values beyond a
possible physical realization, while the main part of the WPD can be
well described by a Gaussian. According to the reasons given above, we
deliberately want to test the EN theory in this situation.

For this test we first calculate numerically the WPD asymptotic
behavior using the EN theory (the numerical method used for solving
the EN equations is described in Appendix~\ref{appx:Numerics}) and fit
the form predicted in Tab.~\ref{tab:table1}, $p(w)\sim
|w|^{-\nu}\exp(-B|w|)$, to these data. Then we compare the fitted
coefficients $\nu$ and $B$ with the results predicted in
\cite{Ryabov2013}. Results for the protocol
\begin{equation}
\lambda(t) = \frac{k}{1+rt}
\label{eq:driving}
\end{equation}
with $k=1$, $r=1$, and $t_f=1$ are shown in
Fig.~\ref{fig:log_harmonic_EQ} for three different $g$ values.  They
strongly suggest that the EN theory is correct even for slow
driving. The parameters $\nu$ and $B$, determined from fits in the $w$
interval $[-350,w_+]$, are shown in the insets as functions of the
upper bound $w_+$ of the fit interval. As can be seen from these
graphs, $\nu$ and $B$ converge to the predicted values with decreasing
$w_+$. It becomes clear also that the convergence to the exact
asymptotic functional form occurs at very large negative work values,
where $p(w)$ attains extremely small values of order $10^{-300}$.
Irrespective of this slow convergence to the asymptotic functional
form, it should be noted that the data from the EN theory are very
close to the exact solution of the complete WPD in the whole $w$
interval shown in Fig.~\ref{fig:log_harmonic_EQ}.

In the following we now assume the EN theory to be valid. The FF
conjecture is tested against it and against BD simulations, which have
been carried out by numerical integration of the Langevin equation
(\ref{eq:langevin}) with the Heun algorithm \cite{Saito1993}. WPD
determined from the BD simulations have been calculated from typically
$10^8$ simulated trajectories.
 
%%%%%%%%%%%%%%%%%%%%%%%%%%%%%%%%%%%%%%%%%%%%%%%%%
%%%%%%%%%%%%%%%%%%%%%%%%%%%%%%%%%%%%%%%%%%%%%%%%%
\section{FF conjecture versus EN theory for equilibrium initial condition}
\label{sec:equilibrium}
%%%%%%%%%%%%%%%%%%%%%%%%%%%%%%%%%%%%%%%%%%%%%%%%%
%%%%%%%%%%%%%%%%%%%%%%%%%%%%%%%%%%%%%%%%%%%%%%%%%

%%%%%%%%%%%%%%%%%%%%%%%%%%%%%%%%%%%%%%%%%%%%%%%%%
\begin{figure}[tb]
\resizebox{1.0\linewidth}{!}{%
  \includegraphics{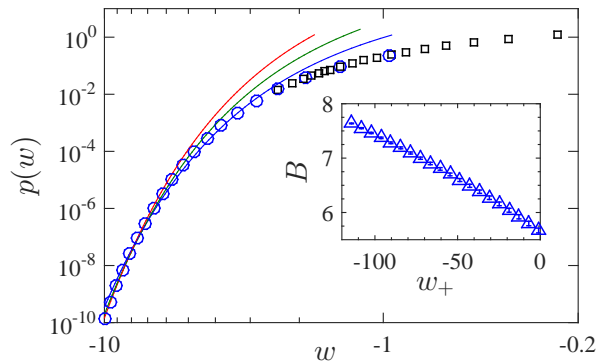}
}
\caption{Asymptotic behavior of the WPD for the potential
  (\ref{eq:uh}) as obtained from the EN theory (circles) and BD
  simulation (squares) for the protocol in Eq.~(\ref{eq:driving}) with
  parameters $k=t_f=1$ and $r = 10$. The three lines are fits of the
  FF conjecture $p(w) \sim |w|^{-3/4}\exp(- B |w|)$ to the EN data for
  different fitting intervals $[-120,w_+]$ with $w_+=-7.0$, $-4.1$, and
  $-1.1$.  The inset shows the fitting parameter $B$ as a function of
  the upper limit $w_+$ of the fit interval. The errors bars given in the inset, which are smaller
    than or comparable to the size of the symbols, mark 95\%
    confidence bounds of fitting parameters. The free energy difference
	reads $\Delta F\approx -0.60$.}
\label{fig:quartic_EQ}
\end{figure}
%%%%%%%%%%%%%%%%%%%%%%%%%%%%%%%%%%%%%%%%%%%%%%%%%

As mentioned in the Introduction, an analytical solution of the
differential equations of the EN theory can be obtained for the
V-shaped potential, which for any monotonic protocol $\lambda(t)$
gives the WPD asymptotic behavior. If $d\lambda(t)/dt>0$, $p(w)$ has
support only for $w>0$, and $p(w)\sim \exp(-Bw)$ for $w\to+\infty$,
while for $d\lambda(t)/dt<0$, $p(w)$ has support only for $w<0$, and
$p(w)\sim \exp\left(|B| w\right)$ for $w\to-\infty$, where
$B=\beta\lambda_0/(\lambda_f-\lambda_0)$.  This functional form is
predicted by the FF conjecture, where, quite surprisingly, in this
case also the parameter $B$ is correctly given.  We have
tested these results against BD simulations (data not shown).

Further tests of the FF conjecture have been performed by us for
various potentials, the potentials (\ref{eq:uh}) and (\ref{eq:us})
given below, the double-well potential $x^4 + \lambda(t)x^2$, various
potentials of the form $\lambda(t)|x|^n$, and the potential
$\lambda(t)\log(1+x^2)$. These tests were carried out for different
protocols, the one given in Eq.~(\ref{eq:driving}), the linear
protocol $\lambda(t)=k+rt$ and the exponential protocol
$\lambda(t)=k\exp(-r t)$. Our findings suggest that for ``hard
potentials'', with stronger confinement than the harmonic one, the FF
conjecture fails in general, and gives, as expected, only a good
approximate description for sufficiently fast driving.

For ``soft potentials'', with weaker confinement than the harmonic
one, there is evidence that the FF conjecture always predicts
correctly the main $|w|\to \infty$ asymptotics of the WPD (exponential
part), and in some cases even the pre-exponential factor. This is true
for all examples in Tab.~\ref{tab:table1}, which refer either to soft
potentials or the harmonic one. It is possible, however, that the
pre-exponential factor is not correctly captured by the FF conjecture.

In the following we exemplify these findings for one hard and one soft
potential, namely
\begin{equation}\label{eq:uh}
U(x,\lambda(t)) = \lambda(t) x^4\,,
\end{equation}
and 
\begin{equation}\label{eq:us}
U(x,\lambda(t)) = \lambda(t)|x| + |x|^{3/2}\,,
\end{equation}
with the protocol $\lambda(t)$ given in Eq.~(\ref{eq:driving}).

Figure~\ref{fig:quartic_EQ} shows the results of the EN theory
(circles) and BD simulations (squares) for the hard potential
(\ref{eq:uh}), which coincide in the asymptotic regime $w \ll \Delta
F$.  The FF conjecture predicts the asymptotic behavior of WPD $p(w)
\sim |w|^{-3/4}\exp(- B |w|)$. Fits of this functional form to the EN
data for three fitting intervals with different upper bounds $w_+$ are
depicted by the lines. As the $B$ factor, shown in the inset of
Fig.~\ref{fig:quartic_EQ}, shows no tendency to converge to a constant
with decreasing upper limit $w_+$ of the fit interval $[-120,w_+]$
(see also the non-overlapping of the three lines in the main part of
the figure), we conclude that the FF conjecture fails in this case.

\begin{figure}[tb]
\resizebox{1.0\linewidth}{!}{%
\includegraphics{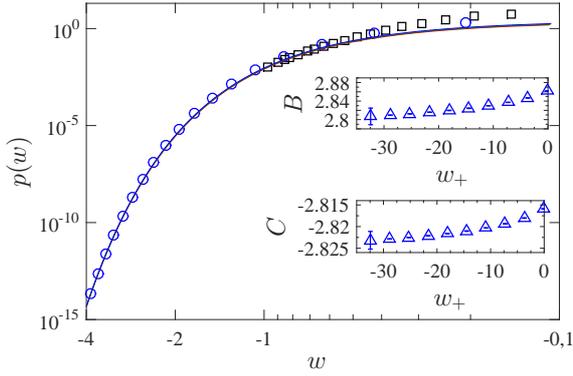}
}
\caption{Asymptotic behavior of the WPD for the potential
  (\ref{eq:us}) as obtained from the EN theory (circles) and BD
  simulation (squares) for the protocol in Eq.~(\ref{eq:driving}) with
  parameters $k=r=t_f=1$. The three overlapping lines are fits of the FF
  conjecture $p(w) \sim \exp(-Bw - C|w|^{3/2})$ to the EN data for
  different fitting intervals [$-36,w_+$] with upper bounds
  $w_+=-25.3$, $-14.5$, and $-3.7$. The insets show the fitting
  parameters $B$ and $C$ as functions of $w_+$. The errors bars given in the insets, which are smaller
    than or comparable to the size of the symbols, mark 95\%
    confidence bounds of fitting parameters.  The free energy difference
  reads $\Delta F\approx -0.24$.}
\label{fig:Cauchy_EQ}
\end{figure}

Figure~\ref{fig:Cauchy_EQ} shows the results of the EN theory
(circles) and BD simulations (squares) for the soft potential
(\ref{eq:us}), which again coincide in the asymptotic regime $w \ll
\Delta F$. The FF conjecture predicts the asymptotic behavior $p(w)
\sim \exp(-Bw - C|w|^{3/2})$. Fits of this functional form to the EN
data for three fitting intervals $[-36,w_+]$ with different upper
bounds $w_+$ are depicted by the lines. The $B$ and $C$ factors from
the fits clearly tend to constants, see also the overlapping of the
three lines in the main part of the figure. These lines match well
also the simulated data. Accordingly, the analysis provides evidence
that the FF conjecture in this case predicts correctly the main
$|w|\to \infty$ asymptotics of WPD.  In the asymptotic formula from
the FF conjecture, the pre-exponential factor equals one. In order to
test also this prediction, we have further fitted the generalized form
$p(w) \sim |w|^D\exp(-Bw - C|w|^{3/2})$ to the numerical data. In this
case we found that the $D$ factor from the corresponding fit shows no
tendency to converge to zero with increasing upper bound $w_+$ [see Fig.~\ref{fig:prefactor}(a)]. We hence can conclude that the FF conjecture fails to
predict the pre-exponential factor in this case.

%%%%%%%%%%%%%%%%%%%%%%%%%%%%%%%%%%%%%%%%%%%%%%%%%
%%%%%%%%%%%%%%%%%%%%%%%%%%%%%%%%%%%%%%%%%%%%%%%%%

%%%%%%%%%%%%%%%%%%%%%%%%%%%%%%%%%%%%%%%%%%%%%%%%%
%%%%%%%%%%%%%%%%%%%%%%%%%%%%%%%%%%%%%%%%%%%%%%%%%
\section{FF conjecture versus EN theory for nonequilibrium initial condition}
\label{sec:nonequilibrium}
%%%%%%%%%%%%%%%%%%%%%%%%%%%%%%%%%%%%%%%%%%%%%%%%%
%%%%%%%%%%%%%%%%%%%%%%%%%%%%%%%%%%%%%%%%%%%%%%%%%

The generalization of the EN theory to a \emph{nonequilibrium} initial
distribution $\rho_0(x)$ is presented in the
Appendix~\ref{appx:ENnonequilibrium}. We compared the predicted
asymptotic behavior from the generalized EN theory for various initial
distributions against extensive BD simulations and found always a good
agreement, both for hard and soft potentials in the sense introduced
above (comparison of the confinement with respect to the harmonic
potential).

Based on the findings in the previous section, an extension of the FF
conjecture to \emph{nonequilibrium} initial distributions is
considered for soft potentials only. A straightforward extension is to
replace the $\rho_{\rm eq}(x)$ in Eq.~(\ref{eq:equilibrium}) by
$\rho_0(x)$, i.e.,\
\begin{equation}
p(w) \underset{\rm fun}{=}
\int dx\, \delta \left[w- \Delta U(x)  \right] \rho_{0}(x)
\,, \quad w \!\to\! {\pm}\infty\,.
\label{eq:neq_largereq}
\end{equation}
When testing this ansatz, we indeed found functional forms predicted
by Eq.~(\ref{eq:neq_largereq}) to fit the data of the EN
theory. However, this support of Eq.~(\ref{eq:neq_largereq}) was only
obtained, if $\rho_0(x)$ is ``wide'' compared to the equilibrium
distribution $\rho_{\rm eq}(x)$ in the sense that $\lim_{|x| \to
  \infty} \rho_0(x)/\rho_{\rm eq}(x) = \infty$.  If $\rho_0(x)$ is
``narrow'' compared to the equilibrium distribution $\rho_{\rm
  eq}(x)$, meaning $\lim_{|x| \to \infty} \rho_0(x)/\rho_{\rm
  eq}(x)<\infty$, our tests showed that Eq.~(\ref{eq:equilibrium})
rather than Eq.~(\ref{eq:neq_largereq}) fits the results of the EN
theory. From a consideration of protocols with infinitely fast
driving, this can be understood by studying a one-step protocol,
where, when starting from the distribution $\rho_{0}(x)$, the
potential is suddenly changed by $\Delta U(x)$ at some time instant
$t_\star>0$ in the interval $]0,t_f]$. If the propagator of the
    dynamics of the position is known, one can calculate the WPD
    analytically for this one-step protocol by replacing $\rho_0(x)$
    by the position distribution $\rho(x,t_\star)$ at time $t_\star$
    in Eq.~(\ref{eq:neq_largereq}).  Investigating potentials with
    explicit solutions of the underlying Smoluchowski equation, we
    indeed found that for wide $\rho_0(x)$ the results were in
    agreement with Eq.~(\ref{eq:neq_largereq}), while for narrow
    $\rho_0(x)$ the results were in agreement with
    Eq.~(\ref{eq:equilibrium}).

To sum up, the FF conjecture for the non-equilibrium initial
distribution suggests that for wide initial conditions the weights of
the trajectories corresponding to large $|w|$ values are determined
solely by the initial distribution and the potential. On the other
hand, for narrow initial conditions the FF conjecture suggests that
these weights are determined rather by the evolved distribution at
some time $t_{\star}>0$, which in turn yields the same WPD asymptotic
behavior as the equilibrium distribution.

%%%%%%%%%%%%%%%%%%%%%%%%%%%%%%%%%%%%%%%%%%%%%%%%%
\begin{figure}[tb]
\resizebox{1.0\linewidth}{!}{%
\includegraphics{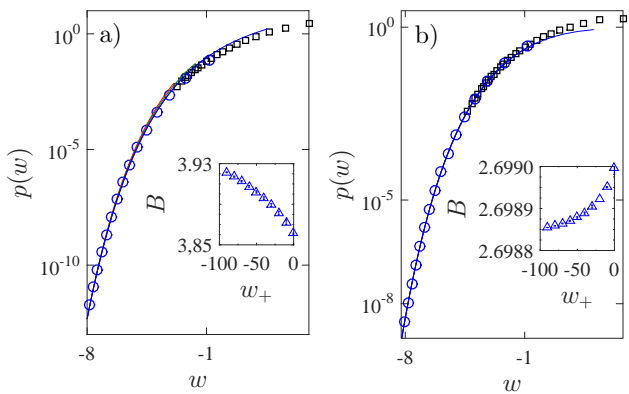}
}
\caption{Asymptotic behavior of the WPD for the breathing parabola
  $U(x,\lambda(t))=\lambda(t)x^2$ for a nonequilibrium initial
  distribution $\rho_0(x)$ and for the protocol in
  Eq.~(\ref{eq:driving}) with $k=r=t_f=1$.  The circles refer to the
  data calculated from the EN theory and the squares refer to data
  obtained from the BD simulations.  In (a) the initial distribution
  is narrow [$\rho_0(x) \propto \exp(-x^4)$] and in (b) the initial
  distribution is wide [$\rho_0(x) \propto x^2\exp(-x^2)$].  The three overlapping
  lines are fits of the FF conjectures, $p(w) \sim |w|^{-1/2}\exp(- B
  |w|)$ in (a) and $p(w) \sim |w|^{1/2}\exp(- B |w|)$ in (b), to the
  EN data for different fitting intervals $[-100,w_+]$ with the upper
  bounds $w_+= -7.0$, $-4.1$, and $-1.1$. The insets show the fitting
  parameters $B$ as functions of $w_+$. The errors bars given in the insets, which are smaller
    than or comparable to the size of the symbols, mark 95\%
    confidence bounds of fitting parameters. The mean works are (a) $\left<
  w\right> = -0.26$, (b) $\left< w\right> = -0.49$.}
\label{fig:noneq}
\end{figure}
%%%%%%%%%%%%%%%%%%%%%%%%%%%%%%%%%%%%%%%%%%%%%%%%%

\begin{figure}[tb]
\resizebox{1.0\linewidth}{!}{%
  \includegraphics{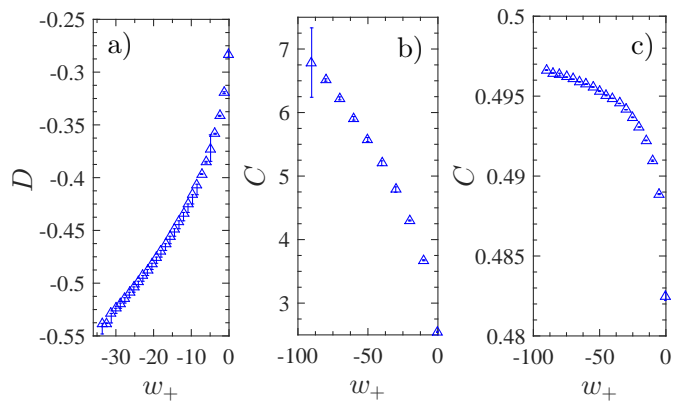}
}
\caption{Test of pre-exponential factors predicted from the FF conjecture. In (a) we used the potential (\ref{eq:us}) and the equilibrium initial condition $\rho_{\rm eq}(x)$. The panel shows the fitting parameter $D$ from the asymptotic form $p(w) \sim |w|^D\exp(-Bw - C|w|^{3/2})$ as a function of the upper bound $w_+$ of the fitting interval [$-36,w_+$]. Panels (b) and (c) correspond to the potential $U(x,\lambda(t))=\lambda(t)x^2$ and a nonequilibrium initial distribution. In (b) the initial distribution is narrow [$\rho_0(x) \propto \exp(-x^4)$] and in (c) the initial distribution is wide [$\rho_0(x) \propto x^2\exp(-x^2)$]. The panels show the fitting parameter $C$ from the asymptotic form $p(w) \sim |w|^{C}\exp(- B |w|)$ as a function of the upper bound $w_+$ of the fitting interval [$-100,w_+$]. The errors bars given in the figures, which are smaller than or comparable to the size of the symbols, mark 95\% confidence bounds of fitting parameters.}
\label{fig:prefactor}
\end{figure}

To exemplify our findings, we present results here for the case of the
breathing parabola $U(x,\lambda(t))=\lambda(t)x^2$, using the protocol
$\lambda(t)$ in Eq.~(\ref{eq:driving}) with $k=r=t_f=1$ and two
nonequilibrium initial distributions $\rho_0(x)\propto\exp(-x^4)$
(narrow) and $\rho_0(x)\propto x^2\exp(-x^2)$ (wide).  As can be seen
from Fig.~\ref{fig:noneq}, in the asymptotic regime $w \ll \left< w
\right>$, where $\left< w \right>$ denotes the mean work done on the
system, the results of the generalized EN theory (circles) agree well
with the simulated data (squares) for both initial distributions.  For
the narrow $\rho_0(x)$, the FF conjecture predicts the asymptotic
behavior $p(w) \sim |w|^{-1/2}\exp(- B |w|)$, and for the wide
$\rho_0(x)$, it predicts $p(w) \sim |w|^{1/2}\exp(- B |w|)$.  Fits of
these functional forms to the EN data for three fitting intervals with
different upper bounds $w_+$ are depicted by the lines.  The $B$
factors from the fits, shown in the insets, tend to approach constants
for decreasing $w_+$.  In particular the data in
Fig.~\ref{fig:noneq}(b) for the wide $\rho_0(x)$ have almost reached a
plateau value [see the $B$ scale in the inset in comparison to the one
  in Fig.~\ref{fig:noneq}(a)].  Note also that the three fitting lines
in the main parts of Fig.~\ref{fig:noneq}(a) and (b) can hardly be
distinguished on the scale of the graphs, which gives further support
that the FF conjecture predicts correctly the main $|w|\to\infty$
asymptotics of WPD. Also in this case we have tested the predicted
pre-exponential factors $|w|^{\pm 1/2}$ by fitting the generalized
formula $p(w) \sim |w|^{C}\exp(- B |w|)$ to the numerical data.  The results suggest that in the case of narrow initial condition the factor $C$ from the corresponding fit shows no tendency to converge to $-1/2$ [see Fig.~\ref{fig:prefactor}(b)]. On the other hand, for the wide initial condition the factor $C$ from the corresponding fit clearly converges to $1/2$ with decreasing $w_+$ [see Fig.~\ref{fig:prefactor}(c)]. We hence can conclude that the FF conjecture fails to predict the pre-exponential factor for the narrow initial condition, while its prediction is precise for the wide initial condition.

%%%%%%%%%%%%%%%%%%%%%%%%%%%%%%%%%%%%%%%%%%%%%%%%%
%%%%%%%%%%%%%%%%%%%%%%%%%%%%%%%%%%%%%%%%%%%%%%%%%

To summarize, the main $|w|\to \infty$ asymptotic behavior predicted
by the generalized EN theory agrees well with simulated data, and
there is good evidence that the FF conjecture for nonequilibrium
initial distributions holds for soft potentials. The pre-exponential
behavior predicted by the FF conjecture may be incorrect.

%%%%%%%%%%%%%%%%%%%%%%%%%%%%%%%%%%%%%%%%%%%%%%%%%
%%%%%%%%%%%%%%%%%%%%%%%%%%%%%%%%%%%%%%%%%%%%%%%%%
\section{Summary and perspectives}
%%%%%%%%%%%%%%%%%%%%%%%%%%%%%%%%%%%%%%%%%%%%%%%%%
%%%%%%%%%%%%%%%%%%%%%%%%%%%%%%%%%%%%%%%%%%%%%%%%%

We have shown that the functional form of the WPD asymptotic behavior
for overdamped Brownian motion in confining potentials can be obtained
often from a simple conjecture, in which it is assumed that the
respective form is independent of the driving velocity.  This
conjecture was motivated by the fact that it is valid for all
potentials and protocols, where analytical solutions for the WPD have
been obtained in the literature. In fact, these potentials are either
the harmonic one or they belong to what we classified as soft
potentials, with weaker confinement than the harmonic potential.

Tests of the FF conjecture have been performed against the predictions
of the EN theory.  With respect to that theory, (i) we gave further
evidence of its validity, in particular in the case of slow driving,
by a comparison of results with the exact WPD for the
logarithmic-harmonic potential, (ii) we have obtained an analytical
solution for the V-potential in addition to the formerly known
parabolic potentials, and (iii) we have extended the theory to the
case of nonequilibrium initial distributions of the particle position.

The tests of the FF conjecture indeed showed good agreement for soft
potentials. For hard potentials, with stronger confinement than the
harmonic potential, our tests strongly indicate that the FF conjecture
is not valid both for equilibrium and nonequilibrium distributions of
the initial particle position.  In the nonequilibrium case for soft potentials, 
the tests were most convincing for wide initial distributions, which decay
slower than the equilibrium one.  However, also for narrow initial
distributions, decaying faster than the equilibrium one, our results
indicate the FF conjecture to hold true. All our new findings were
also checked by BD simulations whenever it was possible.

It is surprising that for all analytically known cases the FF
conjecture is valid. Whether this is a mere coincidence or has a
deeper reasoning needs to be clarified. To this end, theoretical
insight should be gained, why the FF conjecture fails for hard
potentials.

With respect to applications, the FF conjecture can be useful to guess
reasonable functional forms for the WPD asymptotic behavior, which, as
pointed out in the Introduction, are needed, when measured WPD data
have to be extended to the tail regime for use of the JE. A quite
general form of the WPD asymptotic behavior was suggested in
\cite{Palassini2011} and, based on it, a ``Jarzynski estimator'' for
better determination of free energy differences in experiments was
proposed. In cases where it is possible to guess functional forms of
the driving potential, the FF conjecture can be more specific and
utilized to develop improved Jarzynski estimators, or to choose a most
appropriate one among the possible driving potentials. If an
appropriate potential is found, a modeling of the process by BD
simulations and an application of the EN theory can be used to
identify the parameters in this potential.

In our work here we considered overdamped Brownian motion in one
dimension under monotonic protocols.  It would be interesting to check
in future studies, whether extensions of the FF conjecture (and also
of the EN theory) to non-monotonic protocols, to Brownian motion in
higher dimensions, and to the underdamped case are possible. In fact,
the FF conjecture has been seen to be valid also for a specific case
of underdamped diffusion in a breathing parabola \cite{Kwon2013}.

%%%%%%%%%%%%%%%%%%%%%%%%%%%%%%%%%%%%%%%%%%%%%%%%%%%%%%%%%%%%%%%%%%%%%%%%%%%%%%%%%%%%%%%%%%%%%%%%%%%%%%%%%%%%%%%%%%
\begin{acknowledgement}
Support by the Ministry of Education of the Czech Republic (project
no.\ 7AMB14DE003), and by the Deutsche Akademische Austauschdienst
(DAAD, project no.\ 57066299) is gratefully acknowledged.
\end{acknowledgement}

\subsection*{Author contribution statement}
V.H. proposed to study the FF conjecture and together with A.R. tested
its validity using the EN theory. V.H. also derived the results for
the V-potential and generalized the EN theory for the case of
nonequilibrium initial distributions of the particle position.
D.L. performed all computer simulations used in the
manuscript. P.M. supervised the work at every stage, proposed to
perform the tests of the FF conjecture using the EN theory and to test
the EN theory against the results obtained for the log-harmonic
potential. P.Ch. supervised the work at every stage and proposed to
study work distributions for driven Brownian motion.

\appendix

%%%%%%%%%%%%%%%%%%%%%%%%%%%%%%%%%%%%%%%%%%%%%%%%%
%%%%%%%%%%%%%%%%%%%%%%%%%%%%%%%%%%%%%%%%%%%%%%%%%
\section{EN theory}
%%%%%%%%%%%%%%%%%%%%%%%%%%%%%%%%%%%%%%%%%%%%%%%%%
%%%%%%%%%%%%%%%%%%%%%%%%%%%%%%%%%%%%%%%%%%%%%%%%%
\label{appx:ENmethod}

According to \cite{Engel2009,Nickelsen2011}, the WPD asymptotic
behavior for overdamped Brownian motion of a particle in a
time-dependent confining potential $U(x,\lambda(t))$ [Eq.~(\ref{eq:langevin})],
for the initial particle position in equilibrium with distribution
$\rho_\mathrm{eq}(x) = \exp[-\beta U(x,0)]/Z_\mathrm{eq}$, reads
\begin{equation}
p(w) = 
\frac{\sqrt{2}{\cal N}}{Z_\mathrm{eq}}
\frac{\exp(-\beta S)}{\sqrt{(\det A) \langle\dot{V}'| A^{-1} |\dot{V}'\rangle}}
(1+O(\beta^{-1}))\,,
\label{eq:tails_Engel}
\end{equation}
where
\begin{eqnarray}
{\cal N} &=& \exp\left(\frac{1}{2} \int_0^{t_f}dt\,V''_t\right)\,,\label{eq:N}\\
\det A &=& 2 (V_{t_f}'' \chi_{t_f} + \dot{\chi}_{t_f})\,,\label{eq:detA}\\
\langle\dot{V}'| A^{-1} |\dot{V}'\rangle &=&\int_0^{t_f}dt\,\psi_t \dot{V}_t'\,,\label{eq:vav}
\end{eqnarray}
and
\begin{equation}\label{eq:action}
S = V_0 + \int_0^{t_f}dt\,\left[\frac{1}{4}(\dot{y_t}+V')^2 + \frac{q}{2}\dot{V}\right] - \frac{q}{2}w\,,
\end{equation}
where $y_t$ stands for the most probable trajectory which yields the
work value $w$ ($\dot y_t = dy_t/dt$). In all the above formulas the
mobility $\mu$ [cf.\ Eq.~(\ref{eq:langevin})] was set to unity, and we
have abbreviated the time dependence by the lower index. The function
$V_t$ is defined as $V_t = U(y_t,\lambda(t))$, $\dot{V}_t =
\partial{U}(x,\lambda(t))/\partial t|_{x=y_t}$, $V'_t =
\partial{U}(x,\lambda(t))/\partial x|_{x=y_t}$, and analogously for higher
derivatives.

The ``optimal trajectory'' $y_t$ minimizing the action $S$ and the
auxiliary functions $\psi_t$ and $\chi_t$ are obtained from the
following system of three second-order ordinary differential equations
\begin{subequations}
\label{eq:aux-func}
\begin{eqnarray}
\ddot y_t &=&  (q-1)\dot{V}'_t + \dot{V}'_t\dot{V}''_t\,,\quad\\
\ddot\psi_t &=& (V''^2_t +V'_t V'''_t - (1-q)\dot{V}''_t)\psi_t - \dot{V}'_t\,,\label{eq:psi_boundary}\\
\ddot \chi_t &=& (V''^2_t +V'_t V'''_t - (1-q)\dot{V}''_t)\chi_t\,,\label{eq:chi_boundary}
\end{eqnarray}
\end{subequations}
with the boundary conditions
\begin{subequations}
\label{eq:boundary}
\begin{align}
\dot y_0 &= V'_0\,, & \dot y_{t_f}&= - V'_{t_f}\,,\\
\dot\psi_0 &= V''_0\psi_0\,, & \dot\psi_{t_f}&= - V''_{t_f} \psi_{t_f}\,,\\
\chi_0 &= 1\,, & \dot\chi_0&= V''_0\,.
\end{align}
\end{subequations}
The constraint
\begin{equation}\label{eq:w-constraint}
w=\int_0^{t_f}dt\,\dot V_t
\end{equation}
fixes the $q$ value in Eqs.~(\ref{eq:action}),
(\ref{eq:psi_boundary}), and (\ref{eq:chi_boundary}). 

%%%%%%%%%%%%%%%%%%%%%%%%%%%%%%%%%%%%%%%%%%%%%%%%%
%%%%%%%%%%%%%%%%%%%%%%%%%%%%%%%%%%%%%%%%%%%%%%%%%
\section{EN theory applied to V-potential}
%%%%%%%%%%%%%%%%%%%%%%%%%%%%%%%%%%%%%%%%%%%%%%%%%
%%%%%%%%%%%%%%%%%%%%%%%%%%%%%%%%%%%%%%%%%%%%%%%%%
\label{appx:ENabsvalue}

For the potential $U(x,\lambda(t)) = \lambda(t)|x|$ the right hand
side of Eq.~(\ref{eq:tails_Engel}) can be calculated analytically. As
we show below, for large enough values of $|w|$ the optimal
trajectories $y_t$ never crosses the origin. Since we are interested
in the tail behavior $|w|\to \infty$, this means that the
non-analytical nature of $|x|$ at the origin does not influence the
asymptotic behavior of the WPD and in the
Eqs.~(\ref{eq:N})-(\ref{eq:boundary}), one can safely assume that
$V_t''=V_t'''=0$ etc.\ for all $x$. Then the differential equations
for the functions $\psi_t$ and $\chi_t$, which determine the
pre-exponential factor in (\ref{eq:tails_Engel}) become independent of
$y_t$ and thus of $w$ also.

Due to the symmetry of the problem, for each work value $w$ two
optimal trajectories $y_t$ symmetric relative to the origin exist. In
the following, we will solve the equations presented in the preceding
section assuming that $y_t>0$. The trajectories $y_t<0$ can be
incorporated simply by including the factor 2 in the prefactor of the
WPD. The optimal trajectory $y_t$ follows from the equation
\begin{equation*}
\ddot{y}_t = (q-1)\dot{\lambda}_t\,, 
\end{equation*}
which can be easily solved. The two unknown integration constants
together with the auxiliary variable $q$ are determined from the
formulas
\begin{equation*}
\quad \dot{y}_0 = \lambda_0\,, \quad \dot{y}_{t_f} = - \lambda_{t_f}\,, \quad \int_0^{t_f}dt\,\dot{V}_t = w\,.
\end{equation*}
The solution is (for monotonic driving, where $\lambda_{t_f}\ne \lambda_0$)
\begin{equation}
\label{eq:optimal_trajectory_absx}
y_t = f_t + \frac{1}{\lambda_{t_f} - \lambda_0}w\,,\quad q = - \frac{2 \lambda_0}{\lambda_{t_f} - \lambda_0}\,,
\end{equation}
where $f_t$ is a function independent of $w$. Due to the work definition (\ref{eq:work}), the term $w/(\lambda_{t_f} - \lambda_0)$ is for a monotonic driving always positive. The function $f_t$ is finite and thus there always exists some $|w|$ large enough that the whole function $y_t$ is positive for an arbitrary $t$. The action (\ref{eq:action}) can be rewritten as \cite{Nickelsen2011}
\begin{multline}\label{eq:actionII}
S = 
-\frac{w}{2} + \frac{1}{2}(V_{t_f}+V_0) -\frac{1}{4}(y_{t_f}V'_{t_f} + y_0 V'_{0})\\
+ \frac{1}{4}\int_0^{t_f}dt\,V'_t(V'_t-y_t V''_t) + \frac{1-q}{4}\int_0^{t_f} dt\,y_t\dot{V}'_t\,.
\end{multline}
After inserting the optimal trajectory (\ref{eq:optimal_trajectory_absx}) into this formula one obtains
\begin{align*}
S &= -\frac{w}{2} + \frac{1}{4}(\lambda_{t_f}y_{t_f}+\lambda_0 y_0)
+ \frac{1}{4}\int_0^{t_f}dt\,\lambda_t^2 + \frac{1-q}{4}w \\
&= h_{t_f}+\frac{\lambda_0}{\lambda_{t_f} - \lambda_0}w\,,
\end{align*}
where $h_t$ is a another function independent of $w$.  Accordingly,
the WPD fulfills the asymptotic relation
\begin{equation}
p(w) \sim \left\{\begin{array}{ll}
\exp\left(-\frac{\beta\lambda_0}{\lambda_{t_f} - \lambda_0} w\right), \; w\to+\infty\,, & \;\dot\lambda_{t}>0\,,\\
\exp\left(+\frac{\beta\lambda_0}{\lambda_{0} - \lambda_{t_f}} w\right), \; w\to-\infty\,, & \;\dot\lambda_{t}<0\,.
\end{array}\right.
\label{eq:WPD_abs_val}
\end{equation} 
%with prefactors $A_\pm$ independent of $w$. These can in principle also be determined with the EN theory, but we have not carried out the corresponding calculation.

%%%%%%%%%%%%%%%%%%%%%%%%%%%%%%%%%%%%%%%%%%%%%%%%%
%%%%%%%%%%%%%%%%%%%%%%%%%%%%%%%%%%%%%%%%%%%%%%%%%
\section{EN theory for nonequilibrated  initial condition}
%%%%%%%%%%%%%%%%%%%%%%%%%%%%%%%%%%%%%%%%%%%%%%%%%
%%%%%%%%%%%%%%%%%%%%%%%%%%%%%%%%%%%%%%%%%%%%%%%%%
\label{appx:ENnonequilibrium}

The EN theory \cite{Engel2009} can be generalized to a nonequilibrium
initial condition $\rho_0(x) = \exp[-\beta \Gamma(x)]/Z_0$ of the
particle position. The derivation is straightforward and we present
only the results here.

The asymptotic form (\ref{eq:tails_Engel}) remains valid, but the
action is modified,
\begin{equation}\label{eq:s-noeq}
S = \Gamma + \int_0^{t_f}dt\,\left[\frac{1}{4}(\dot{y_t}+V')^2 + \frac{q}{2}\dot{V}\right] - \frac{q}{2}w\,,
\end{equation}
where $\Gamma = \Gamma(y_0)$.  Also the auxiliary functions $y_t$,
$\psi_t$ and $\chi_t$ for determining $S$ from Eq.~(\ref{eq:s-noeq})
and $\det A$ and $\langle\dot{V}'| A^{-1} |\dot{V}'\rangle$ from
Eqs.~(\ref{eq:detA}) and (\ref{eq:vav}), respectively, become
modified. Their evolution equations (\ref{eq:aux-func}) remain the
same, but the boundary conditions now are
\begin{subequations}
\label{eq:boundaryNEQ}
\begin{align}
\dot y_0 &= 2\Gamma' - V'_0\,, & \dot y_{t_f} &= - V'_{t_f}\,,\\
\dot\psi_0 &= (2\Gamma'' - V''_0)\psi_0\,, & \dot\psi_{t_f} &= - V''_{t_f} \psi_{t_f}\,,\\
\chi_0 &= 1\,, & \dot\chi_0 &= 2\Gamma'' - V''_0\,.
\end{align}
\end{subequations}

%%%%%%%%%%%%%%%%%%%%%%%%%%%%%%%%%%%%%%%%%%%%%%%%%
%%%%%%%%%%%%%%%%%%%%%%%%%%%%%%%%%%%%%%%%%%%%%%%%%
\section{Numerical procedure used for solving EN equations}
%%%%%%%%%%%%%%%%%%%%%%%%%%%%%%%%%%%%%%%%%%%%%%%%%
%%%%%%%%%%%%%%%%%%%%%%%%%%%%%%%%%%%%%%%%%%%%%%%%%
\label{appx:Numerics}

In order to solve the EN equations (Appendices~\ref{appx:ENmethod} and
\ref{appx:ENnonequilibrium}), we have adopted the numerical procedure
``bvp4c'' implemented in MATLAB and previously used by Nickelsen
\cite{Nickelsen-thesis}. Bvp4c is a finite difference code, which
solves two-point boundary value problems for ordinary differential
equations with possible further unknown parameters. It implements the
three-stage Lobatto IIIa formula \cite{Butcher1987}. This is a collocation formula and
the collocation polynomial provides a continuous solution with
continuous first derivative that is fourth order accurate uniformly in
[$a,b$]. Mesh selection and error control are based on the residual of
the continuous solution.

In the code, an initial guess of the solution is used to iteratively
find the correct solution. In our implementation we guessed the
solution for some small absolute value of $w$, and use the
corresponding solution as the initial guess for a new, slightly
changed, work value. The corresponding solution is then again used as
the guess for another work value and so on until a given final value
of $w$ is reached. Problems arise when the solutions of EN equations
for two close work values differ significantly. Then the numerical
procedure collapses, because it is not able to converge from a distant
initial guess to the solution. In our analysis we have encountered
such numerical problems when using very soft potentials such as
$U(x,\lambda(t)) = \lambda(t) \ln (1+x^2)$.

%%%%%%%%%%%%%%%%%%%%%%%%%%%%%%%%%%%%%%%%%%%%%%%%%%%%%%%%%%%%%%%%%%%%%%%%%%%%%%%%%%%%%%%%%%%%%%%%%%%%%%%%%%%%%%%%%%
% BibTeX users please use
\bibliographystyle{epj}
\bibliography{bibliography}

\begin{thebibliography}{21}
\bibitem{comm:units}
With the potential are generally associated characteristic length scales, which
  could be used to define an appropriate length unit in the problem. Because we
  will consider different potentials, we refrain to introduce them here. One
  could alternatively introduce a laboratory length scale $\xi$ as length unit.
  Then the time unit would be given by $\mu k_\mathrm{B}T/\xi^2$.
\end{thebibliography}
%
% Non-BibTeX users please use
%\begin{thebibliography}{}
%
% and use \bibitem to create references.
%
%\bibitem{RefJ}
% Format for Journal Reference
%Author, Journal \textbf{Volume}, (year) page numbers.
% Format for books
%\bibitem{RefB}
%Author, \textit{Book title} (Publisher, place year) page numbers
% etc
%\end{thebibliography}

\end{document}